\documentclass[usenatbib]{mn2e}
\usepackage{apjfonts,amsfonts,amsmath,amssymb,bm,ctable,verbatim}
\usepackage{pdflscape}

\newcommand{\Alf}{{Alfv\'en}}
\newcommand{\bhat}{\hat{\bf b}}
\newcommand{\HII}{H~{\sc ii}}
\newcommand{\HI}{H~{\sc i}}
\newcommand{\acknowledgments}[1]{\begin{small}\section*{Acknowledgments}\end{small}{\noindent #1}\vspace{5pt}}
\newcommand{\datastatement}[1]{\begin{small}\section*{Data Availability Statement}\end{small}{\noindent #1}\vspace{5pt}}

\title[CR Scattering due to Intermittent Structures]{Galactic Cosmic-ray Scattering due to Intermittent Structures}

\author[Butsky et al.]{
\parbox[t]{\textwidth}{
Iryna S.~Butsky$^{1,2}$\thanks{ibutsky@stanford.edu}\thanks{NASA Hubble Fellow}, 
Philip F.~Hopkins$^1$, 
Philipp Kempski$^3$,
Sam B.~Ponnada$^1$,
Eliot Quataert$^3$,
Jonathan Squire$^4$
}\vspace*{4pt} \\
$^1$ TAPIR, Mailcode 350-17, California Institute of Technology, Pasadena, CA 91125, USA \\
$^2$ Kavli Institute for Particle Astrophysics and Cosmology and Department of Physics, Stanford University, Stanford, CA 94305, USA \\
$^3$ Department of Astrophysical Sciences, Princeton University, Princeton, NJ 08544, USA \\
$^4$ Physics Department, University of Otago, Dunedin 9010, New Zealand
}

\date{}
\begin{document}
\maketitle

\begin{abstract}
Cosmic rays (CRs) with energies $\ll$\,TeV comprise a significant component of the interstellar medium (ISM). Major uncertainties in CR behavior on observable scales (much larger than CR gyroradii) stem from how magnetic fluctuations scatter CRs in pitch angle. Traditional first-principles models, which assume these magnetic fluctuations are weak and uniformly scatter CRs in a homogeneous ISM, struggle to reproduce basic observables such as the dependence of CR residence times and scattering rates on rigidity. We therefore explore a new category of ``patchy'' CR scattering models, wherein CRs are predominantly scattered by intermittent strong scattering structures with small volume-filling factors. These models produce the observed rigidity dependence with a simple size distribution constraint, such that larger scattering structures are rarer but can scatter a wider range of CR energies. To reproduce the empirically-inferred CR scattering rates, the mean free path between scattering structures must be $\ell_{\rm mfp}\sim10\,{\rm pc}$ at GeV energies.  We derive constraints on the sizes, internal properties, mass/volume-filling factors, and the number density any such structures would need to be both physically and observationally consistent. We consider a range of candidate structures, both large-scale (e.g.\ \HII\, regions) and small-scale (e.g.\ intermittent turbulent structures, perhaps even associated with radio plasma scattering) and show that while many macroscopic candidates can be immediately ruled out as the primary CR scattering sites, many smaller structures remain viable and merit further theoretical study. We discuss future observational constraints that could test these models.
\end{abstract}


\section{Introduction}
\label{sec:intro}

\begin{figure*}
\centering
\includegraphics[width=0.96\textwidth]{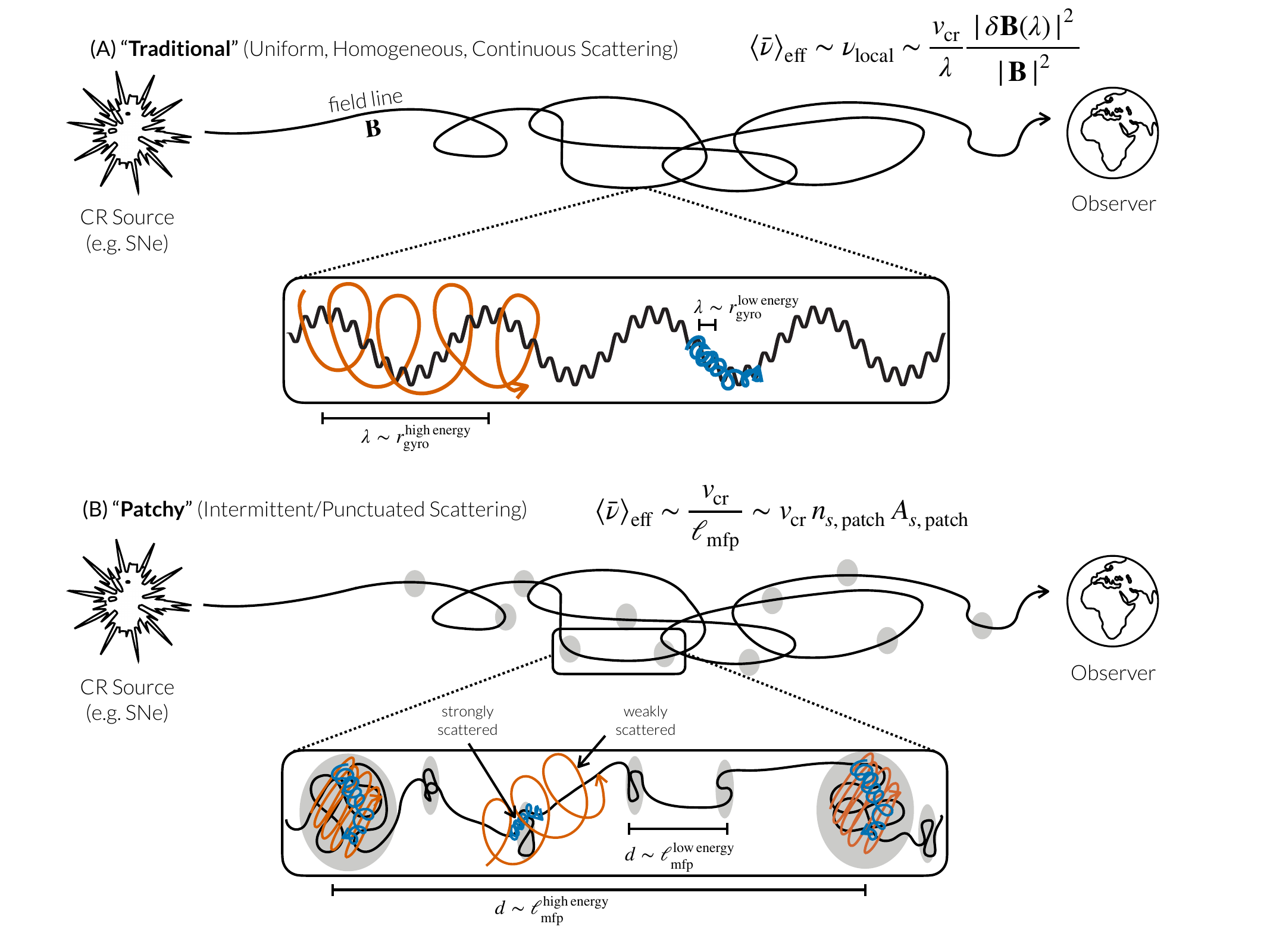}
\caption{Cartoon illustrating the difference between traditional models of sub-TeV CR pitch-angle scattering in the ISM (Section \ref{sec:problem}; {\em top}) and the ``patchy'' models proposed here (Section \ref{sec:patchy}; {\em bottom}). Low-energy CRs are accelerated at some sites (e.g.\ supernovae) and confined to gyro orbits with small radii, $r_{\rm gyro}$, and some pitch angle, $\mu$, along magnetic field lines, ${\bf B}$, through the ISM, until they are observed. 
``Traditional'' models (Section \ref{sec:problem}) assume CR pitch angles are scattered constantly by ubiquitous (volume-filling factor $\sim1$) small-amplitude ($|\delta{\bf B}|\ll |{\bf B}|$) magnetic fluctuations (with wavelength $\lambda$ here) uniformly distributed though a medium that is homogeneous up to scales of the CR deflection length $\ell_{\rm mfp} \sim 10\,{\rm pc}\,R_{\rm GV}^{\delta_s}$, giving rise to some effective mean scattering rate $\langle \bar{\nu}_{\rm eff} \rangle$ shown. Reproducing the observed dependence of CR scattering rates and residence times on rigidity therefore requires a specific ratio of median/typical volume-filling magnetic fluctuations of different scales $\lambda$.
The ``patchy'' model (Section \ref{sec:patchy}; {\em bottom}) assumes that CR scattering is dominated by strong scattering in intermittent ``patches'' or structures (gray ovals), which have a small volume-filling factor (with size $\ell_{S}$, and number density $n_{s}$). Larger patches, which have an effectively larger ``CR scattering optical depth'' and scatter both low- and high-energy CRs, are rarer than smaller patches, which only scatter low-energy CRs. An appropriate distribution of smaller and larger patches can therefore produce the observed dependence of CR scattering on rigidity. } 
\label{fig:cartoon} 
\end{figure*}

Low-energy ($\sim$ MeV - TeV) CRs contain most of the CR energy and pressure \citep[e.g.][]{Cummings:2016} and as such impact astrochemistry, ISM plasma physics, and galaxy evolution on all scales, from protoplanetary disks \citep[e.g.][]{Turner:2009, Padovani:2018}, to molecular clouds \citep[e.g.][]{Herbst:1973, Goldsmith:1978, Padovani:2009} and the diffuse ISM \citep[e.g.][]{Heintz:2020, Simpson:2023}, to far-reaching galactic outflows \citep[e.g.][]{Ipavich:1975,Hopkins:2021_outflows, Quataert:2022} that shape the circumgalactic and intergalactic medium \citep[CGM/IGM; e.g.][]{Guo:2008, Butsky:2018, Ji:2020}. CRs in this energy range are also crucial probes of fundamental high-energy astrophysics, astroparticle physics, and ISM plasma physics on scales that cannot be observationally resolved \citep{Zwei13,zweibel:cr.feedback.review,2018AdSpR..62.2731A,2019PrPNP.10903710K, Mollerach:2018, Gabici:2019}. Despite their importance, low-energy CRs remain poorly understood.  

The vast majority of the literature studying low-energy CR propagation and dynamics has focused on simple, phenomenological prescriptions for the effective CR transport rates within the ISM, typically parameterized with an effective diffusion coefficient $\kappa_{\rm eff}$, or streaming speed ${\rm v}_{\rm st, eff}$.  Using a variety of methods, classic studies constrained these coefficients by comparing detailed models of CR propagation in a Galactic background to observed CR spectra in the Solar system (including many CR species and a range of CR energies, ratios of primary-to-secondary CRs, radioactive and isotopic abundances, as well as the CR anisotropy on the sky), and/or to diffuse $\gamma$-ray observations  \citep{strong:2001.galprop,vladimirov:cr.highegy.diff,blasi:cr.propagation.constraints,gaggero:2015.cr.diffusion.coefficient,2016ApJ...819...54G,2016ApJ...824...16J,cummings:2016.voyager.1.cr.spectra,2016PhRvD..94l3019K,evoli:dragon2.cr.prop}. The ``effective'' coefficients inferred by such studies represent, by definition, some weighted average in the ISM between CR sources (e.g.\ supernova remnants in the Milky Way) and the Solar system. Together, the existing observations still only constrain the CR scattering physics in ISM conditions broadly similar to the Solar neighborhood. Since phenomenological models do not explain how such scattering rates actually arise or break many of the degeneracies between CR propagation models, it is by no means clear how to extrapolate their findings to distinct environments (e.g.\ the Galactic center, CGM, or IGM), galaxy types (e.g.\ dwarf, starburst, or high-redshift), or CR acceleration cites (e.g.\ AGN and quasars). Moreover, different choices for such extrapolation can lead to orders-of-magnitude different results in predicted galaxy properties \citep[see references above and][]{hopkins:2020.cr.transport.model.fx.galform,butsky:2020.cr.fx.thermal.instab.cgm,butsky:2023.cr.kappa.lower.limits.cgm}. 

In order to predict how CRs propagate on macro scales in different galactic environments, we first need to understand CR physics on micro scales. The gyroradii, $r_{\rm gyro}$, of low-energy CRs are extremely small compared to the ``macroscopic'' scales of the ISM/CGM (e.g.\ $r_{\rm gyro} \sim 0.1\,$au, for CRs at the $\sim\,$GeV peak of the spectrum in the diffuse ISM). This means that CRs cannot simply escape their acceleration sites around supernovae at bulk speeds ${\rm v}_{\rm cr} =\beta_{\rm cr} c$, but instead travel along magnetic field lines on gyro orbits with a characteristic gyro frequency $\Omega \sim {\rm v}_{\rm cr}/r_{\rm gyro}$, and pitch angle $\mu \equiv \hat{\bf v}_{\rm cr} \cdot \bhat$ relative to the magnetic field direction $\bhat \equiv {\bf B}/|{\bf B}|$. When CRs encounter magnetic field fluctuations, $\delta {\bf B}$, they are scattered in pitch angle, producing some effective pitch-angle scattering rate, $\nu_{\rm eff}$. When averaged over large spatial and temporal scales, $\nu_{\rm eff}$ leads to bulk CR transport that can be parameterized by some effective diffusion coefficient $\kappa_{\rm eff} \sim {\rm v}_{\rm cr}^{2}/\nu_{\rm eff}$ and/or streaming speed ${\rm v}_{\rm st,\,eff} \sim \kappa_{\rm eff} |\nabla P_{\rm cr}|/P_{\rm cr}$. As we summarize in Section \ref{sec:problem}, traditional scattering models typically assume that the magnetic fluctuations that scatter CRs are weak ($|\delta {\bf B}| / |{\bf B}| \ll 1$) and uniformly distributed throughout the volume-filling ISM, but differ from each other in their proposed origin of the magnetic fluctuations.

Constraining CR scattering theories is extremely difficult since it is simply not possible to directly resolve the relevant gyroresonant scales in either ISM observations or in numerical simulations that also include macroscale ISM processes. Even idealized particle-in-cell type simulations of CR scattering in an ISM ``patch'' that is only modestly larger than the CR gyroradii (so obviously unable to span the full dynamic range of conditions) have only just become possible in the last few years \citep{bai:2019.cr.pic.streaming,bai:2022.streaming.instab.sims,holcolmb.spitkovsky:saturation.gri.sims,plotnikov:2021.cr.mhd.pic.sims.streaming.ion.neutral.strong.damping.deconfines,bambic:2021.mhd.pic.transport.inhomogeneous.ionization.effective.confinement.very.different.from.length.corr.of.turb.much.similar.to.cr.fluid.dynamics.models,ji:2021.cr.mhd.pic.dust.sims,ji:2021.mhd.pic.rsol}. Because of these challenges, existing CR scattering models remain wildly uncertain, even in the ``typical'' ISM. For example, when applied in galaxy simulations, state-of-the-art scattering models that are calibrated to reproduce existing observations predict CR scattering rates that differ by as much as ten orders of magnitude (at $\sim\,$GeV energies) in the ISM and predict qualitatively different scalings with properties like magnetic field strength and turbulence \citep{hopkins:cr.transport.constraints.from.galaxies}. Additionally, multiple recent studies have pointed out that existing scattering models struggle to even \textit{qualitatively} capture the correct dependence of CR scattering rate on rigidity at sub-TeV energies \citep{Hopkins:2022,Kempski:2022}.

In this paper, we are therefore motivated to propose a novel picture for CR scattering, which may resolve some of these challenges. In Section \ref{sec:problem}, we summarize some of the central challenges facing ``conventional'' models in the recent literature. In Section \ref{sec:patchy}, we present a new theoretical framework for ``patchy'' CR scattering, which is qualitatively distinct from traditional, continuous scattering models and derive a number of criteria any such model must obey to reproduce observations. We discuss a variety of candidate scattering structures, both macroscopic and small-scale in Section \ref{sec:examples}. We show which candidate scattering structures can be immediately ruled out  and discuss the potential connections to intermittent structures in turbulence. In Section \ref{sec:obs} we discuss the observational implications for these model categories. We summarize and conclude our results in Section \ref{sec:summary}.

\section{The Problem with Simple, Homogeneous Theories of Cosmic-ray Scattering}
\label{sec:problem}

Empirical constraints on CR propagation in the Galaxy typically infer an effective diffusivity of the form  $D_{xx} \sim \kappa_{\rm eff} = D_{0}\,\beta_{\rm cr}\,(R_{\rm cr}/R_{\rm cr,\,0})^{\delta_{\rm s}}$, equivalent to an angle-averaged CR pitch-angle scattering rate, 
\begin{equation}
\label{eqn:scattering_empirical}
\langle \bar{\nu} \rangle_{\rm eff} \sim \bar{\nu}_0 \beta_{\rm cr} R^{-\delta_s}_{\rm GV},
\end{equation}
where $\beta_{\rm cr} = {\rm v}_{\rm cr} / c$ is the CR velocity, $R_{\rm GV}$ is the CR rigidity in GV, and typical values of the fit parameters correspond to  $\bar{\nu}_0 \sim 10^{-9}\, {\rm s}^{-1}$, and $ 0.5 \lesssim\, \delta_s \lesssim 0.7$ \citep[e.g.][and references in Section \ref{sec:intro}]{delaTorre:2021.dragon2.methods.new.model.comparison}. This is roughly equivalent to an isotropic diffusion coefficient $\kappa_{\rm eff} \sim 10^{29}\,{\rm cm}^2\,{\rm s}^{-1}$, or effective streaming velocity ${\rm v}_{\rm st,\,eff} \sim 300\,{\rm km}\, {\rm s}^{-1}$, assuming a gradient scale length of 1 kpc. The effective CR scattering rate translates into a characteristic mean free (or deflection) time, $t_{\rm mfp} \sim 1/\langle \bar{\nu} \rangle_{\rm eff}$, or mean free (or deflection) path, $\ell_{\rm mfp} \sim {\rm v}_{\rm cr}/\langle \bar{\nu} \rangle_{\rm eff} \sim (c\, / \bar{\nu}_{0})\,R_{\rm GV}^{\delta_s}$,  between $\mathcal{O}(1)$ deflections in $\mu$.

Regardless of the details, an extremely robust observational result is $\delta_s > 0$: the CR residence times and scattering rates (which scale $\propto \langle \bar{\nu} \rangle_{\rm eff} \propto R_{\rm GV}^{-\delta_s}$) must {\em decrease} with increasing CR rigidity at energies $\sim\,$GeV-TeV, in order to reproduce any of the observed trends, e.g. secondary-to-primary or radioactive isotopic species ratios.

Historical theories of CR scattering (heuristically illustrated in Fig.~\ref{fig:cartoon}) generally assume that the observed CR pitch-angle scattering is the cumulative result of a large number of uncorrelated, small perturbations to $\mu$ generated by encounters with a very large number of independent, small magnetic field fluctuations, $\delta {\bf B}$. These theories also assume that the magnetic field fluctuations occur throughout an ISM that is {\em statistically homogeneous} on spatial scales from the gyro scale up to, or larger than, the CR deflection length. This is a questionable starting assumption considering the CR deflection length $\ell_{\rm mfp} \sim 10\,{\rm pc}\,R_{\rm GV}^{\delta_s}$ (as large as hundreds of pc for $\sim$\,TeV CRs) is larger than the size scales of much ISM structure. 
Nonetheless, integrating over an ensemble of perturbations, this leads to the classic predicted scattering rate $\langle \bar{\nu} \rangle_{\rm eff} \sim ({\rm v}_{\rm cr}/\lambda)\,|\delta {\bf B}(\lambda)|^{2}/|{\bf B}|^{2}$ \citep[e.g.][]{Voelk1973}, where $|\delta {\bf B}(\lambda)| \ll |{\bf B}|$ represents some typical (e.g.\ root-mean-square) fluctuation amplitude with wavelength $\lambda$.\footnote{$\lambda$ is often taken to be the gyroresonant wavelength $\lambda\sim r_{\rm gyro}$ since that will usually dominate if there is an undamped spectrum of fluctuations, but it can also represent scattering by larger-wavelength modes ($\lambda \gtrsim r_{\rm gyro}$) via e.g.\ transit-time damping \citep[see e.g.][]{yan.lazarian.02, Malyshkin:2002}.} This applies equally well to both ``self-confinement'' theories (in which $\delta{\bf B}$ is sourced at $\lambda \sim r_{\rm gyro}$ by CR streaming instabilities; \citealt{Kulsrud:1969,wentzel.1969.streaming.instability,Skilling:1975a}) and ``extrinsic turbulence'' theories (in which $\delta {\bf B}$ is sourced by a turbulent cascade from vastly-larger ISM driving scales; \citealt{jokipii:1966.cr.propagation.random.bfield,Voelk1973}).

However, as discussed extensively in \citet{Hopkins:2022} and \citet{Kempski:2022}, the most commonly-invoked self-confinement and extrinsic turbulence theories based on the above assumptions do not reproduce the locally-observed CR spectra at sub-TeV energies. For example, putting in typical ISM values for the relevant parameters gives orders-of-magnitude different normalization ($\bar{\nu}_{0}$) from that observed \cite[a point already made in][]{chandran00,yan.lazarian.02,hopkins:cr.transport.constraints.from.galaxies,chan:2018.cosmicray.fire.gammaray,fornieri:2021.comparing.et.models.data.et.only.few.hundred.gv}. However, even allowing for arbitrary re-normalization, the CR spectra predicted by traditional CR scattering models will not have the correct shape if one assumes typical scaling parameters \citep[also noted in][]{yan.lazarian.04:cr.scattering.fast.modes,fornieri:2021.comparing.et.models.data.et.only.few.hundred.gv}. Most importantly, these models predict $\delta_s \le 0$, i.e.\ longer CR residence times for higher-energy CRs (opposite the observed behavior). 

The problem with assuming CR scattering is ``continuous'', is that the only way to reproduce the observed dependence $\langle \bar{\nu} \rangle_{\rm eff} \propto R_{\rm GV}^{-\delta_s}$ is to invoke some connection between the dominant wavelength of scattering modes and $R_{\rm GV}$ (for example, assuming gyroresonant scattering $\lambda \sim r_{\rm gyro} \propto R_{\rm GV}$), {\em and} a specific power-law spectrum of fluctuations $\delta {\bf B}(\lambda)$.\footnote{For example, if $\lambda \propto R_{\rm GV}^{\alpha_{\lambda}}$ and $|\delta{\bf B}(\lambda)| \propto \lambda^{\alpha_{B}}$, we require $\alpha_{\lambda}(1-2\,\alpha_{B})\approx \delta$.} But in self-confinement theory, the only stable steady-state solution is one where all CRs either stream at approximately the MHD \Alf\ speed, or free-stream (unconfined) at $c$, in either case, clearly independent of rigidity ($\delta_s =0$). Any solutions out of equilibrium ``collapse'' to these states on a very rapid timescale ($\lesssim$\,Myr) -- an issue that has been noted going back at least to \citet{skilling:1971.cr.diffusion}. In extrinsic turbulence models, it is common to make the phenomenological comparison to Kolmogorov or ``Kraichnan''-type spectra $\delta{\bf B}(\lambda)\propto \lambda^{1/4-1/3}$, which appear at first to give reasonable estimates for $\delta_s$. The problem is that this assumes the turbulence is both undamped and isotropic down to scales at least as small as the gyroresonant wavelength, which cannot be true for sub-TeV CRs, where the gyro scales are much smaller than the \Alf\ and Kolmogorov/damping/dissipation scales of the turbulence. In that regime, turbulence is highly anisotropic, and the parallel structure necessary for efficient CR scattering are suppressed \citep{Goldreich:1995}.\footnote{In the more careful discussions in e.g.\ \citet{Hopkins:2022} and \citet{Kempski:2022}, the distinction between parallel and perpendicular wavenumbers is made more explicitly, and this plays a crucial role in the challenges to traditional extrinsic turbulence theories. We refer to those studies for more detail.} The prediction is then that scattering is necessarily dominated by larger-scale modes, $\lambda\gg r_{\rm gyro}$, which are independent of $r_{\rm gyro}$, so $\delta_s \le 0$, a point also noted in \citet{chandran00,fornieri:2021.comparing.et.models.data.et.only.few.hundred.gv}.

There might be solutions to these problems involving, for example, alternative, volume-filling sources of the modes $\delta {\bf B}$ that differ from standard self-confinement or extrinsic turbulence theories (see \citealt{Hopkins:2022} for discussion). But thus far, there does not appear to be an example of such a model that has actually been shown to reproduce the observed CR spectra.

\section{An Alternative: ``patchy'' scattering by meso-scale structures}
\label{sec:patchy}

Here, we consider a more radical alternative: patchy scattering, in which CR scattering rates are high in discrete ``scattering structures'' and low in-between such structures (illustrated in Fig.~\ref{fig:cartoon}). 

More formally, we drop the assumption of the ``classic'' models in Section \ref{sec:problem} that the fluctuations, $\delta{\bf B}$, from which CRs scatter are homogeneous and uniformly volume filling. In this case, the effective scattering rate, $\langle \bar{\nu} \rangle_{\rm eff}$, inferred from various observational constraints should not be thought of as a uniform, volume-filling rate, but as an average rate of encountering scattering structures. Equivalently, the observationally-inferred effective deflection length, $\ell_{\rm mfp}$, would no longer represent the length over which a sufficient number of small deflections are continuously accrued to change $\mu$ by $\mathcal{O}(1)$, but rather would represent the mean free path between scattering ``patches,''
\begin{equation} 
\ell_{\rm mfp} \sim {\rm v}_{\rm cr} / \langle \bar{\nu} \rangle_{\rm eff} \sim 10\,{\rm pc}\,R_{\rm GV}^{\delta_s},
\end{equation}
Importantly, we will show that this reinterpretation frees us from being forced to assume there is a unique one-to-one correspondence between the measured rigidity dependence, $\delta_s$, and the shape of the power spectrum of magnetic fluctuations on gyro scales, as in Section \ref{sec:problem}.
In the following sections, we discuss the basic constraints that a plausible scattering structure would need to meet.

\subsection{Geometry and General Constraints}
\label{sec:constraints:sizes}

First, consider the basic geometric properties of candidate scattering structures. It is helpful to think of structures in terms of their effective dimensionality, $\mathcal{D}$, equal to the number of dimensions along which the system has a highly elongated axis ratio. Using this definition, $\mathcal{D} = 2$ describes sheet or pancake structures with two long axes ($\ell_L$) and one short axis ($\ell_S$), $\mathcal{D}=1$ describes filamentary (or tube-like) structures with one long axis and two short axes, and $\mathcal{D}=0$ describes spherical (or ``point-like'') structures with all axis lengths of order $\ell_S$. The cross-sectional area of the structure, across a random set of viewing angles, is dominated by the two larger dimensions, $A_s \sim \ell_L^{\mathcal{D}} \ell_S^{2-\mathcal{D}}$, while the relative depth of the structure (as seen by e.g.\ CRs traversing it) is dominated by the short-axis distance, $\ell_S$. From these definitions, the volume of a scattering structure scales as $V_s \sim A_s \ell_S$. 

\subsubsection{Size and Internal Scattering Constraints}
The first, most basic, size constraint is that in order to scatter a CR with some rigidity $R_{\rm GV}$, the structure must be larger than that CR's gyroradius, 
\begin{equation}
    \ell_S \gtrsim r_{\rm gyro} \sim 0.1\, {\rm au}\, R_{\rm GV}/ {B}_{\mu {\rm G}},
\end{equation}
where $B_{\mu {\rm G}}$ is the magnetic field in microGauss. 

Additionally, we assume that there is a local CR pitch-angle scattering rate, $\nu_s$, inside the scattering structures. It is important to distinguish the CR scattering rate within the structures from the ISM-averaged effective scattering rate, $\langle \bar{\nu} \rangle_{\rm eff}$, as by definition $\nu_s \gg \langle \bar{\nu} \rangle_{\rm eff}$. Therefore, the depth of a scattering structure must be large enough so that the scattering time within it ($\sim 1 / \nu_s$) is shorter than the unscattered CR crossing time of that structure ($\sim \ell_S / {\rm v}_{\rm cr}$). In other words, the structure must be large enough so that CRs are reliably scattered as they traverse it, 
\begin{equation}\label{eqn:scattering.depth}
    \ell_S \gtrsim {\rm v}_{\rm cr} / \nu_s. 
\end{equation}
Another way of saying this is that the scattering ``optical depth'' to CRs of some rigidity $R_{\rm GV}$, $\tau_{s} \sim \nu_{s}\,\ell_{S}/{\rm v}_{\rm cr} \gtrsim 1$, must exceed unity in order for CRs of that $R_{\rm GV}$ to be strongly scattered. 

On the other hand, the structure cannot be so large that CRs are effectively trapped inside of it for much longer than their inferred total residence times in the ISM. Assuming there is a large scattering rate inside the structure, the time CRs spend inside of the scattering structure, $t_s$, is set by the effective diffusivity, $\kappa_s \approx {\rm v}_{\rm cr}^2/\nu_s$, giving 
\begin{equation} 
\label{eqn:scattering.time.constraint}    \Delta t_s \sim \ell_S^2 / \kappa_s \sim (\ell_S / {\rm v}_{\rm cr})^2 \nu_s. 
\end{equation}
By definition, if this were longer than the ISM-averaged deflection time $\sim \ell_{\rm mfp}/{\rm v}_{\rm cr}$, then this would dominate the total residence time and exceed the limits above, violating the basic assumptions of our framework. Therefore, we require $\Delta t_s < \ell_{\rm mfp}/{\rm v}_{\rm cr}$, or 
\begin{equation}
\label{eqn:escape.time.constraint}    \ell_S^{2} < {\rm v}_{\rm cr}\ell_{\rm mfp}/\nu_s .
\end{equation}

It is only possible to satisfy both equation \eqref{eqn:scattering.depth} and equation \eqref{eqn:escape.time.constraint} if the depth of the scattering structure is significantly smaller than the mean free path between scattering structures
\begin{equation}
    \ell_S \ll \ell_{\rm mfp}.
\end{equation}
But this is of course implicit in a ``patchy'' scenario.

\subsubsection{Number Densities, Surface Densities, and Mass or Volume-Filling Factors}
\label{sec:constraints:numbers}

The mean free path between patches that can scatter CRs of a given rigidity $R_{\rm GV}$ is set by the cross-sectional area of the structures as well as their relative abundance or ``number density'' $n_s$, as $\ell_{\rm mfp} \sim 1 / (n_s A_s)$. Given the observationally-constrained mean free path for CRs of that rigidity, the scattering structures therefore must have a volume-averaged ISM number density of 
\begin{equation}\label{eqn:ndens}
    n_s \sim 1 / (\ell_{\rm mfp}A_s) \sim 1 / (\ell_{\rm mfp} \ell_L^{\mathcal{D}} \ell_S^{2-\mathcal{D}}).
\end{equation}

Next, consider the typical surface density (or column density; $\Sigma_s$) of a structure, viewed from a random angle, $\Sigma_s \sim M_s/A_s$, in terms of the mass of the structure ($M_s$) and its cross-sectional area ($A_s$). Using the relations above, $\Sigma_s \sim M_s / A_s \sim \bar{\rho}_s \ell_{\rm mfp}$, where $\bar{\rho}_s \sim M_s n_s$ is the volume-averaged mass density of the scattering structures within the ISM as a whole. Assuming that the total mass contained in the scattering structures is some fraction, $f_M$, of the ISM  (with some volume-averaged ISM density $\rho_{\rm ISM} \approx m_{\rm H} n_{\rm ISM}$), we can rewrite the expression for the surface density as
\begin{equation}
    \Sigma_s \sim f_M \rho_{\rm ISM} \ell_{\rm mfp} \sim 5\,f_M \times 10^{-5}\,{\rm g\,cm}^{-2}\,R_{\rm GV}^{\delta_s},
\end{equation}
where $\rho_{\rm ISM} = m_{\rm H}\, n_{\rm ISM}$ is the volume-averaged ISM density, $m_{\rm H}$ is the mass of a hydrogen atom, and $n_{\rm ISM} \approx 1\,{\rm cm}^{-3}$. Assuming that $f_M < 1$, we can place an upper limit on the surface density of scattering structures,
\begin{equation}
    f_M \sim \Sigma_s / (\rho_{\rm ISM} \ell_{\rm mfp}) \sim ({\rho}_{s} /\rho_{\rm ISM})(\ell_S / \ell_{\rm mfp}) < 1,
\end{equation}
where $\rho_s = M_s / V_s$ is the internal density of a scattering structure, i.e.\ $\Sigma_{s} < 5\times10^{-5}\,{\rm g\,cm^{-2}}\,\,R_{\rm GV}^{\delta_s}$ or column density $<3\times10^{19}\,{\rm cm^{-2}}\,R_{\rm GV}^{\delta_s}$. 

Similarly, the volume-filling fraction of the scattering structures, $f_V$, would be
\begin{align}
    f_V \sim V_s n_s \sim \ell_S / \ell_{\rm mfp} < 1, \\
    \ell_{\rm mfp} \sim \ell_S / f_V.
\end{align}

These above quantities ($n_s$, $\Sigma_s$, $f_{M}$, $f_{V}$) depend on CR rigidity either explicitly, or implicitly, through $\ell_{\rm mfp} \propto R_{\rm GV}^{\delta_s}$. This tells us the bulk properties of the scattering patches that scatter lower-energy CRs ($\sim$MeV; e.g. their number densities, column densities, mass and volume-filling factors) will differ from the bulk properties of the patches that scatter higher-energy CRs ($\sim$TeV). Of course, the sizes of the structures ($\ell_S$, $\ell_L$) that scatter CRs of different $R_{\rm GV}$ may also be distinct, as we might naturally expect from the constraints in Section \ref{sec:constraints:sizes}. 

\subsection{Sufficiently Weak Scattering Between Structures}

A key requirement of the patchy scattering model is that CR scattering not be dominated by scattering in the medium between patches. This means that the diffuse/volume-filling ISM cannot have substantial CR scattering from either extrinsic turbulence or from the CR streaming instability (SI). For the extrinsic turbulence case, this is easy to satisfy, as the more detailed calculations in Section \ref{sec:problem} argue that the theoretically-favored extrinsic scattering rates in the warm ISM for sub-TeV CRs are orders of magnitude smaller than the mean observed $\langle \bar{\nu} \rangle_{\rm eff}$. However, one must still avoid runaway growth of the CR SI between patches, which would self-confine CRs to move no faster than the \Alf\ speed, over-confining them especially at higher energies \citep{hopkins:2021.sc.et.models.incompatible.obs}. 

There are a few ways in which the overconfinement problem could be avoided. For example, at energies $\gg$\,GeV, the SI growth rate is proportional to the number density of CRs and so drops rapidly. Thus, for higher energy CRs, the constraint is not so severe, and it may be possible that CRs just around $\sim 1$\,GeV are self-confined while other physics take over between $1-1000$\,GeV \citep[though this may require some fine-tuning; see][]{Kempski:2022}. Additionally, some recent MHD-PIC simulations have argued that SI growth rates may be slower than expected from simple quasi-linear expressions \citep{bai:2019.cr.pic.streaming,holcolmb.spitkovsky:saturation.gri.sims}. 

Alternatively, we can use the weak scattering requirement to place some constraints on the properties of the volume-filling ISM. In steady state, with some large-scale background CR gradient, there would be some net flux of CRs moving away from the galactic center, leading to the growth of SI on some timescales. The patchy scattering model can still hold, so long as the scattering rate due to the saturated SI in the medium between scattering patches is less than the observationally-inferred effective CR scattering rate, $\nu_{\rm SI} < \nu_{\rm eff}$. This is equivalent to requiring that the effective diffusivity due to the SI be larger than the empirically-constrained average diffusion coefficient,
\begin{align}
    \kappa_{\rm SI} \sim \frac{\Gamma_{\rm damp}\,(e_{\rm B}\, c\, r_{\rm gyro} / {\rm v}_{\rm A})}{|\nabla e_{\rm cr}|} > \kappa_{\rm eff} \approx 10^{29}\, {\rm cm}^2\, {\rm s}^{-1},
\end{align}
where $\Gamma_{\rm damp}$ is the effective damping rate of gyroresonant \Alf\ waves, $e_{\rm B} \equiv |{\bf \rm B}|^2/8\pi$ is the magnetic energy density, ${\rm v}_A$ is the \Alf\ speed, and $e_{\rm cr}$ is the CR energy density (equation 7 in \citealt{hopkins:cr.transport.constraints.from.galaxies}). Following the assumptions in \citealt{hopkins:cr.transport.constraints.from.galaxies}, we can turn the above equation into a rough, order-of-magnitude lower limit on the required damping rate to sufficiently suppress the CRSI in the volume-filling ISM, $\Gamma_{\rm damp} \gtrsim 10^{-9}\,{\rm s}^{-1}$, a rate on the higher end of that obtained via simple estimates in the ionized ISM, though potentially reasonable \citep[e.g.][]{Farmer:2004, Zweibel:2017, squire:2021.dust.cr.confinement.damping}. 

\subsection{Key Requirements to Reproduce Observational Scalings and Differences from the Homogeneous Models} 
\label{sec:key}

Provided a potential scattering structure meets all the above criteria, the key point is that by having discrete structures that are not volume filling, we no longer need to impose a specific scattering rate $\nu_s$, or a specific distribution of magnetic field fluctuations $\delta {\bf B}(\lambda)$ inside the structure in order to produce the observed dependence of $\lambda_{\rm mfp}$ or $\langle \bar{\nu}\rangle_{\rm eff}$ on CR rigidity (Section \ref{sec:problem}). Essentially, we have replaced the requirement that the observed scaling $\langle \bar{\nu} \rangle_{\rm eff} \propto R_{\rm GV}^{-\delta}$ reflects a specific power-law scaling of $\delta {\bf B}(\lambda) \propto \lambda^{\alpha_{B}}$, with a different requirement: that the number of scattering structures that can scatter CRs of a given rigidity depends in a specific power-law fashion on that rigidity. 

How plausible is this? First, we consider a simplified scenario in which scattering structures all have some internal scattering rate $\nu_s$, and vary from each other only in size. In this case, the scattering patches would have a wide range of sizes $\ell_{S}$, spanning a range from the smallest to largest gyroradii of the GeV-to-TeV CRs of interest ($\sim 0.1-100\,$au for microGauss fields). Per Section \ref{sec:constraints:sizes}, the patches that are able to scatter CRs of a given rigidity will have $\ell_S > \ell_{S,\,{\rm min}} \sim r_{\rm gyro} \propto R_{\rm GV}$. Our number density constraint then becomes: $n_s(>\ell_S) \propto \ell_S^{\mathcal{D}-2-\delta_s} \ell_L^{-\mathcal{D}}$. Realistically, the scattering rate within the proposed structures will likely have some dependence on CR rigidity in addition to the size of the structure $\nu_s \propto \ell_S^{-\alpha_{S}}\,R_{\rm GV}^{-\alpha_{R}}$. Combined with the requirement that structures can scatter CRs ($\ell_S > {\rm v}_{\rm cr}/\nu_{s}$) and the other scalings from Section \ref{sec:constraints:sizes}-\ref{sec:constraints:numbers}, this gives a more generalized constraint on the number density, $n_{s}(>\ell_S) \propto \ell_S^{-2-\delta_s (1-\alpha_{S})/\alpha_{R} + \mathcal{D}}\,\ell_L^{-\mathcal{D}}$.

In either case, for quasi-3D structures ($\mathcal{D}=0$) or $\ell_{L} \propto \ell_{S}$, reasonable values for $\alpha_S \approx 1$, $\alpha_R \approx 0.5$, and $0.3\lesssim \delta_s \lesssim 0.7$, we obtain $n_s(>\ell_{S}) \propto \ell_S^{-\alpha_{n}}$ with $2 \lesssim \alpha_n \lesssim 3$, which is broadly similar to the distribution of sizes of many classes of objects in the ISM including molecular clouds and \HII\ regions, \HI\ filaments, star clusters, stellar wind termination shocks/bubbles/magnetospheres \citep{guszejnov:universal.scalings}. For sheet-like structures ($\mathcal{D}=2$), the scaling is quite similar to the distribution of shock widths seen in supersonic, isothermal turbulence \citep{squire.hopkins:turb.density.pdf,mocz:2019.markov.turbulence.model}. As advertised, the scaling does not depend sensitively or in the same manner as we discussed in Section \ref{sec:problem} on how $\nu_s$ (or implicitly $\delta {\bf B}$) depends on wavelength $\lambda$.

Thus while it has proven (surprisingly!) challenging to theoretically construct a power spectrum of magnetic fluctuations that satisfies the observational requirements from Section \ref{sec:problem}, it appears, at least in principle, straightforward to conceive of models with a size distribution of ``patches'' that satisfy the relevant requirements (i.e.\ will produce the same observables) without violating any obvious constraints.

\begin{figure}
\centering
\includegraphics[width=0.46\textwidth]{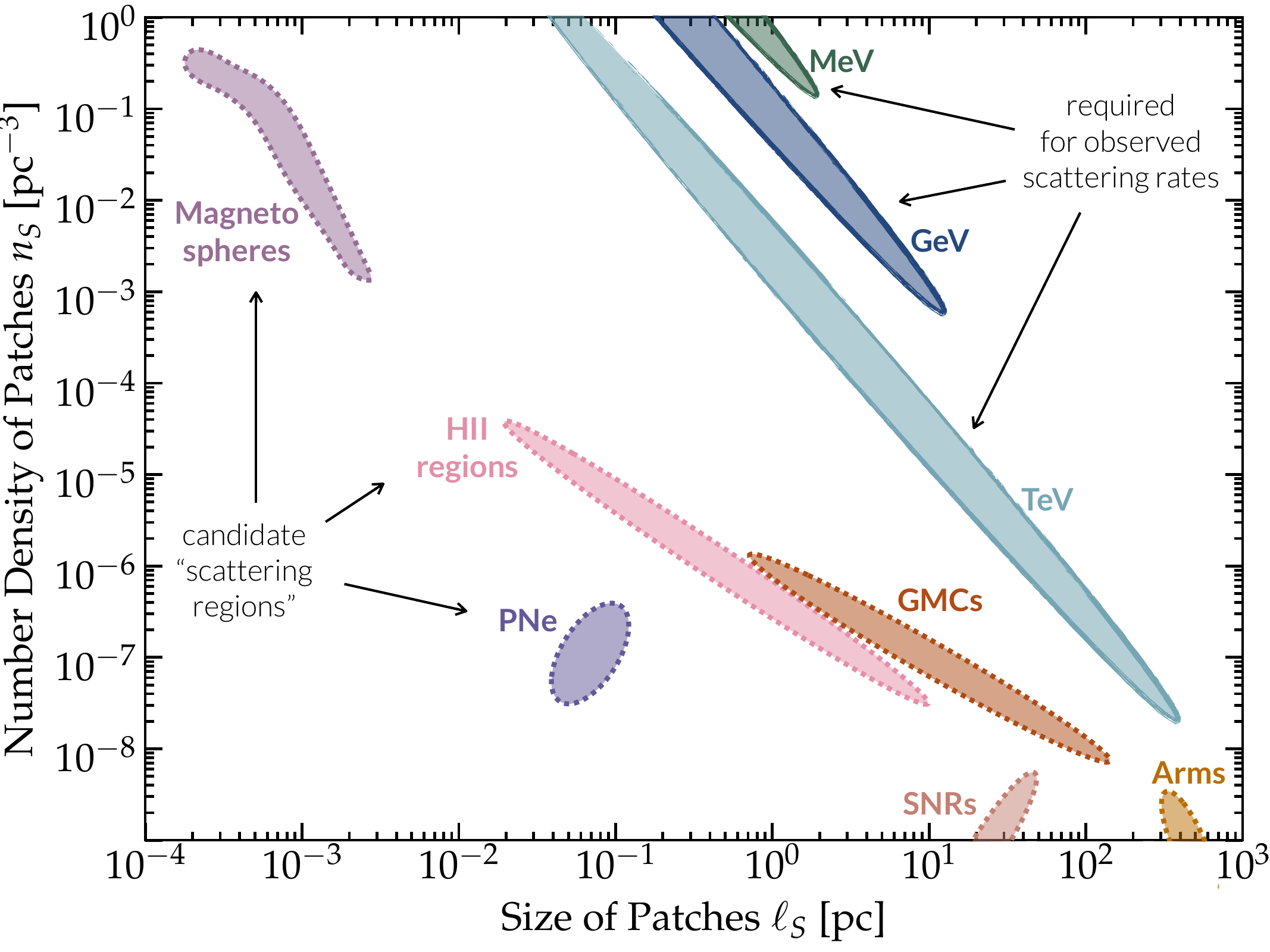}\vspace{-0.2cm}
\caption{Constraints on the size, $\ell_{S}$, and volume number density in the ISM, $n_{\rm s}$, of possible patchy scattering structures (Section \ref{sec:examples:macro}). The blue/green shaded regions with solid boundaries (labeled) show the contours that would produce roughly the correct mean free paths for MeV, GeV, and TeV CRs, obeying all the other constraints in Section \ref{sec:patchy}, for quasi-3D ($\mathcal{D}=0$) structures. The slope of the blue/green regions is set by fixing the CR mean free path, $\ell_{\rm mfp} \sim 1 / (n_s \ell_{S}^{2})$, and the maximum allowed size is set by $\ell_{S} > \ell_{\rm mfp}$. An ideal candidate scattering structure, which could explain CR scattering in the $\sim$ MeV-TeV range, would intercept all of these. The warmer-colored shaded regions with dotted boundaries (labeled) show the approximate location of various known large-scale structures in the ISM including stellar magnetospheres, \HII\ regions, molecular clouds (GMCs), planetary nebulae (PNe), supernova remnants (SNRs), and galactic spiral arms (arms). None of these appear viable: they might scatter CRs, but their abundance is too low to account for most observed CR scattering in the ISM.}
\label{fig:macro_scattering} 
\end{figure}

\section{Example Candidate Structures or Physical Mechanisms}
\label{sec:examples}

We now consider some different physical mechanisms and/or candidate ``scattering structures.'' 

\subsection{Quasi-Static/Coherent and ``Macroscopic'' ISM Structures}
\label{sec:examples:macro}

One possibility is that the ``patches'' of interest could be associated with some known population of quasi-static or coherent ISM structures that are already known to perturb the magnetic field structure on some scale $\ell_{S}$. In evaluating whether such a population is viable as the dominant source of CR scattering, it is helpful to place them on a sort of modified ``Hillas plot'' for CR pitch-angle scattering in the ISM, which we show in Fig.~\ref{fig:macro_scattering}. There, we plot $n_{s}(>\ell_{S})$ versus $\ell_{S}$, for quasi-3D objects ($\mathcal{D}=0$, which reasonably describes all the systems we consider in the plot), and show the allowed regions that produce the observed mean free paths, combining all of the constraints from Section \ref{sec:constraints:sizes}-\ref{sec:constraints:numbers}\footnote{We take the estimated constraints on $\ell_{\rm mfp}$ specifically from \citet{hopkins:cr.multibin.mw.comparison}, though as discussed therein it makes very little difference if we adopt other recent studies' results (compare e.g.\ \citealt{delaTorre:2021.dragon2.methods.new.model.comparison,korsmeier:2022.cr.fitting.update.ams02}), especially given the enormous dynamic range in Fig.~\ref{fig:macro_scattering}. We also assume a diffuse field of a few ${\rm \mu G}$ to estimate $r_{\rm gyro}$, but again changing this by even an order of magnitude does not change our conclusions here.} for CR protons with energies of $\sim$\,MeV, GeV, and TeV. We also place some rough estimates of the range of number densities and sizes of various ``macroscopic'' scattering candidates, including molecular clouds (GMCs), stellar magnetospheres, planetary nebulae (PNe), supernova remnants (SNRs), HII regions, and Galactic spiral arms (rough estimates of number versus size here compiled from \citealt{blitz:2004.gmc.mf.universal,tielens:2005.book,draine:ism.book,walder:magnetic.field.stars.nebulae,anderson:2014.wise.HII.region.compilation,armentrout:2018.hii.census}).

An ideal candidate scattering structure would, in this plot, intersect not just one but all three of the allowed CR ``bands'' without over-predicting the scattering rate for any energy range. Instead, all of the plotted candidates appear clearly ruled out as the dominant source of CR scattering: at a given $\ell_{S}$, $n_{s}$ is orders-of-magnitude too low. In other words, the mean free path between structures is too large, or, alternatively, the maximum ISM-mean scattering rate, $\langle \bar{\nu} \rangle_{\rm eff} \sim {\rm v}_{\rm cr}/\ell_{\rm mfp} \sim {\rm v}_{\rm cr}\,n_{s}\,A_{s} \sim {\rm v}_{\rm cr}\,n_{s}\,\ell_{s}^{2}$, is much smaller than required to account for the observed CR scattering. So while these structures can scatter CRs (we know, in fact, the Heliosphere does so), they cannot produce most of the observed scattering. For this reason structures that have similar sizes but are even rarer in the ISM (e.g.\ Bok globules, pulsar wind nebulae, globular cluster cores, colliding wind binaries) are also immediately ruled out.

Thus while it is conceivable that such a ``macroscopic'' ISM population might exist, we are unable to identify an obvious candidate. Furthermore, the probability that macroscopic structures act as the primary source of CR scattering is reduced due to the fact that macroscopic structures are predominantly found within the galactic disk. Such a disk-centric distribution of scattering candidates contrasts with the inferred CR confinement several kiloparsecs outside of the galactic disk.

\begin{figure}
\centering
\includegraphics[width=0.46\textwidth]{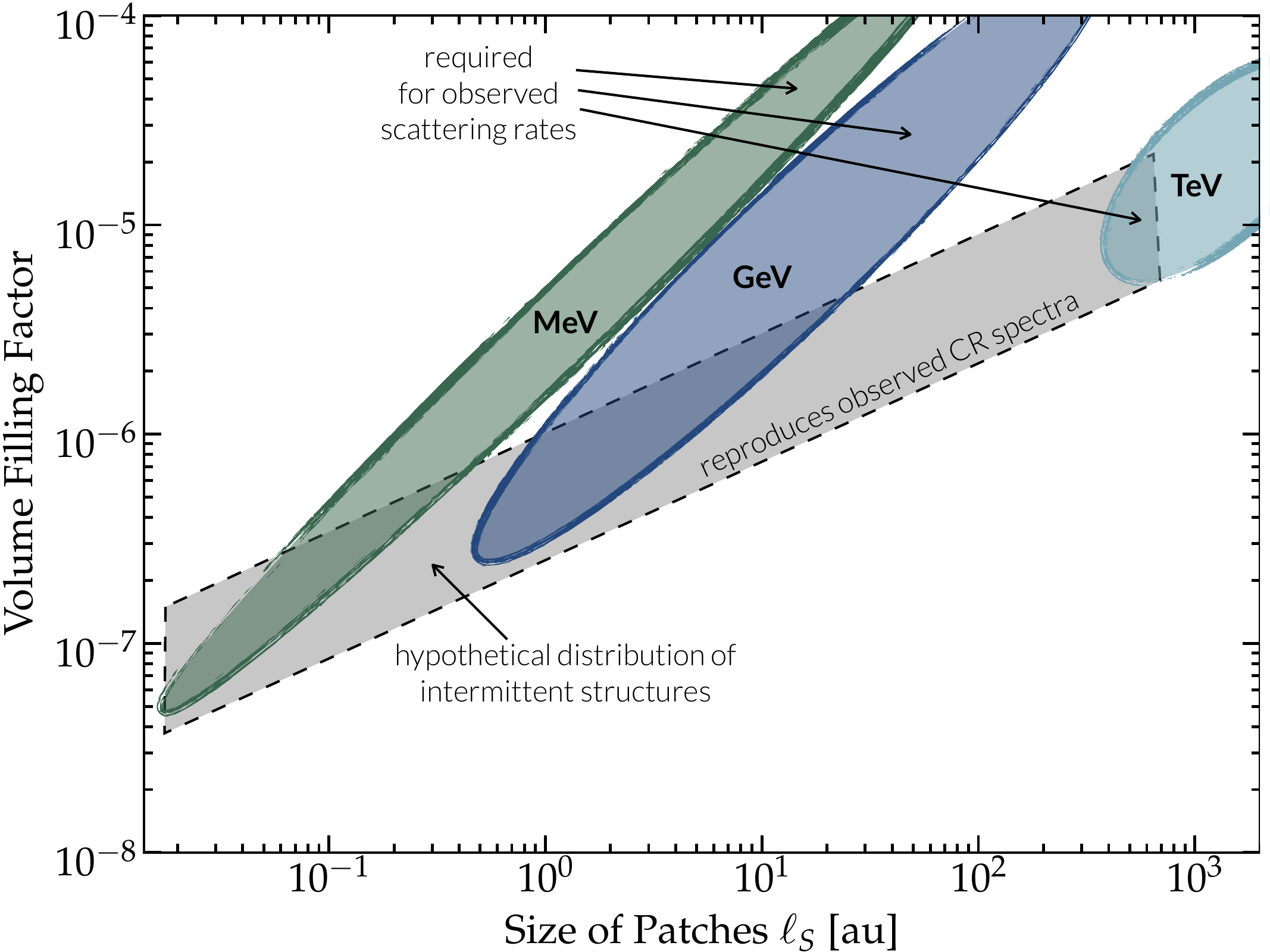}\vspace{-0.2cm}
\caption{Similar to Fig.~\ref{fig:macro_scattering}, we constrain the minor axis, $\ell_S$, and volume-filling factor, $f_V$, required of the patchy scattering model (Section \ref{sec:small.ism}-\ref{sec:small.physics}; this plot is also valid for any dimensionality of structures, $\mathcal{D}=0,\,1,\,2$). We compare the observationally and physically-allowed regions, and provide an example (black shaded region) of a hypothetical model for a distribution of scattering patches, with size $\ell_{S} \sim r_{\rm gyro}$ and $f_V \sim \ell_S/\ell_{\rm mfp} \sim \ell_S^{1/2}$, which would satisfy all of the observational constraints without over-predicting the CR scattering rate or violating any obvious physical or observational limits. This could arise from intermittent structures in turbulence (Section \ref{sec:intermittency}), where a very small volume-filling factor or volumetric probability $P_{V} \sim 10^{-7}-10^{-5}$ of structures with size-scale $\sim \ell_{S} \sim$\,au featuring $\mathcal{O}(1)$ magnetic field fluctuations would be sufficient to explain the observed CR scattering. Interestingly (Section \ref{sec:small.ism}), the size scales here are very similar to those inferred for small-scale ISM structures responsible for radio-wave plasma scattering, and the volume-filling factors $f_{V}$ might be consistent as well, but orders-of-magnitude uncertainties in the observed $f_{V}$ prevent us from reliably placing specific examples on this plot (likewise for some proposed physical mechanisms in Section \ref{sec:small.physics}).}
\label{fig:fvol} 
\end{figure}

\subsection{``Small''-Scale ISM Structure}
\label{sec:small.ism}
An alternative possibility is that the proposed scattering patches could be associated with some small-scale ISM stuctures or magnetic field features.

While Fig.~\ref{fig:macro_scattering} focused on relatively large-scale structures, in Fig.~\ref{fig:fvol}, we repeat the same exercise, but focus on the smaller end of the size range of $\ell_{S}$ (though note the axis ranges do overlap). For small-scale structures, especially where the effective dimensionality may not be known, we find it useful to focus on the volume-filling factor $f_{V} \sim n_{s}\,V_{s} \sim n_{s}\,A_{s}\,\ell_{S} \sim \ell_{S}/\ell_{\rm mfp}$ (Section \ref{sec:constraints:numbers}), rather than $n_{s}$ specifically. This also lets us compress the vertical dynamic range of the plot and factor out the dependence on $\mathcal{D}$, so this plot is valid for sheet-like or filamentary structures, not just quasi-spherical structures. 

In this plot, we also show a hypothetical model that would explain the required rigidity dependence of CR scattering and reproduce Solar-system observables fairly naturally, assuming $\ell_{S} \sim r_{\rm gyro}$ and then calculating $n_{s}(>\ell_{S})$ such that $\ell_{\rm mfp}$ scales with $r_{\rm gyro}$ and therefore $R_{\rm GV}$ as observed (as in Section \ref{sec:key}). Since $f_{V} \sim \ell_{S}/\ell_{\rm mfp}$ we can rearrange to obtain $f_{V} \sim 10^{-7}\,B_{\rm \mu G}^{-\delta_s}\,(\bar{\nu}_{0}/10^{-9}\,{\rm s^{-1}})\,(\ell_{S}/0.2\,{\rm au})^{1-\delta_s} \sim 3\times10^{-7}\,\ell_{S,\,{\rm au}}^{1/2}\,B_{\rm \mu G}^{-1/2}$. 

It is noteworthy that the required volume-filling factors ranging from $f_{V}\sim 10^{-7}-10^{-5}$ for structures with sizes ranging from $\sim 0.01-100\,$au (or $\sim$\,a few $\times10^{-7}$ around $\sim 1\,$au) are intriguingly similar to some estimates of the volume-filling factor of so-called ``tiny-scale atomic structures'' (TSAS; \citealt{heiles:1997.tsas,stanimirovic:2003.tsas.rare,Stanimirovi:2018,mvevoy:2015.tsas.small.volume.filling.factor}) in the ISM as well as the volume-filling factor estimated in {\em some} models of ISM plasma structures causing ``extreme scattering events'' (ESEs; \citealt{Romani:1987,cordes:2001.ese.scattering.structures,Bannister:2016,jow:2023.ese.models}). However, we caution that the ESE filling factor is largely unconstrained and model-dependent \citep{Stanimirovi:2018}. While the sizes ($\ell_{S}$) of TSAS structures are broadly agreed to lie in the range plotted in Fig.~\ref{fig:fvol}, some other observational estimates of their volume-filling factor are as high as $\sim 10^{-2}$, much larger than what is needed to explain CR scattering \citep[see e.g.][]{brogan:2005.high.tsas.covering.factor}.

Note these are categories of ISM structures classified by their effects on radio waves: physical explanations for such structures range widely, but often invoke intermittent turbulent structures, which we discuss below.

\subsection{Connection to Intermittency}
\label{sec:intermittency}

A natural category of candidates for the scattering structures suggested by Fig.~\ref{fig:fvol} is intermittent turbulent structure \citep{Zhdankin:2016, Mallet:2017, Dong:2018}. 
Recent test particle simulations of CRs in intermittent magnetic fields suggest that magnetic structure may enhance CR diffusion, even for a fixed magnetic field power spectrum \citep{Shukurov:2017, Seta:2018}. The anomalous CR diffusion due to magnetic field intermittency has qualitatively similar behavior to the more general approach of modeling CR diffusion in Fokker-Plank-like equations with non-Markovian statistics \citep[e.g.][]{Wilk:1999, Snodin:2016, Zimbardo:2020}. Below, we place constraints on magnetic intermittency in the context of the patchy CR scattering model.

Recall, in the traditional model, we had $\langle \bar{\nu} \rangle_{\rm eff} \sim ({\rm v}_{\rm cr}/\lambda)\,|\delta{\bf B}(\lambda)|^{2}/|{\bf B}|^{2}$, with the assumption that scattering was dominated by ubiquitous (volume-filling factor $f_{V} \sim 1$) but weak ($|\delta{\bf B}| \ll |{\bf B}|$) fluctuations, which must obey specific conditions on their power spectra at scales $\ll 100\,$au in order to reproduce the observed dependence of CR residence time on rigidity. In the patchy model here, we have $\langle \bar{\nu} \rangle_{\rm eff} \sim {\rm v}_{\rm cr}/\ell_{\rm mfp} \sim ({\rm v}_{\rm cr}/\ell_{S})\,f_{V}$ (using $f_{V} \sim \ell_{S}/\ell_{\rm mfp}$ from Section \ref{sec:constraints:numbers}). So, per Fig.~\ref{fig:fvol}, we instead assume that CR scattering is dominated by regions with strong magnetic fluctuations (``patches'') but very low volume-filling factor ($f_{V} \ll 1$). For example, if we were to assume gyroresonant $\lambda \sim \ell_{S} \sim r_{\rm gyro}$, then in order to reproduce the observed $\langle \bar{\nu} \rangle_{\rm eff} \propto R_{\rm GV}^{-\delta_s}$, in the ``traditional'' models we must have $|\delta {\bf B}(\lambda)|^{2} \propto \lambda^{1-\delta_s}$, while in the patchy model we replace this with the requirement $f_{V}(\ell_{S}) \propto \ell_{S}^{1-\delta_s}$. While the latter does not (and indeed cannot, mathematically) reduce the number of observational requirements on the model, it does avoid all of the mathematical and physical challenges to the ``traditional'' models. Specifically, the intermittent scattering model removes the requirement that CR scattering theories produce spectra of the form $|\delta {\bf B}(\lambda)|^{2} \propto \lambda^{1-\delta_s}$ at sub-100\,au scales in the ISM, as reviewed in \citet{Hopkins:2022} and \citet{Kempski:2022}. 

In Fig.~\ref{fig:fvol}, we show an example of a hypothetical successful intermittent scattering model (black shaded region), assuming the size of scattering structures scales with the CR gyroradius, $\ell_S \sim r_{\rm gyro}$ and $\delta_s \sim 1/2$. In order for a scattering patch with a size scale of $\mathcal{O}(1)$ gyroradius to reliably scatter CRs (have a ``scattering optical depth'' of order unity) at that rigidity, this hypothetical model requires a magnetic fluctuation amplitude $\mathcal{O}(|\delta{\bf B}(\ell_{S})| /|{\bf B}|) \sim 1$. So, as heuristically demonstrated in Fig.~\ref{fig:fvol}, a model that features intermittent structures with $\mathcal{O}(|\delta{\bf B}(\ell_{S})| /|{\bf B}|) \sim 1$ on size scales $\ell_{S} \sim 1-1000$\,au, with small volume-filling factor $f_{V} \sim 10^{-6}\,(\ell_{S}/10\,{\rm au})^{1/2}$, would automatically give rise to the ``desired'' (empirically-inferred) CR scattering rates at $\lesssim\,$GeV through $\gtrsim$\,TeV energies. The exact scaling, of course, would also depend on the details of how the magnetic field strength scales with the size of scattering patches \citep[e.g.][]{Lemoine:2023, Kempski:2023}.

We can also reason about the prevalence of intermittent turbulent structures in terms of the shape of the probability distribution function (PDF) of magnetic fluctuations. Consider the volumetric PDF $P_{V}(\delta{\bf B}\,|\,\ell_{S})$ of fluctuations $|\delta {\bf B}|$ with a given size scale/wavelength $\ell_{S}$: to calculate the contribution of different fluctuations with the given $\ell_{S}$ to CR scattering we should integrate over this PDF. If the PDF is Gaussian/normal -- i.e.\ completely non-intermittent -- then the contribution to CR scattering will be dominated by the ``weak'' $\pm 1\sigma$ fluctuations in the core of the PDF, giving rise to the usual scattering rate $\propto \langle |\delta {\bf B}|^{2} \rangle \sim |\delta {\bf B}|_{\rm median}^{2}$. But now, as is commonly parameterized for intermittent systems, consider a PDF with power-law tails in the rare-event (large-$\delta{\bf B}$) regime, $dP_{V}/d\ln{|\delta{\bf B}|} \propto |\delta{\bf B}|^{-\alpha_{P}}$ (i.e.\ $dP_{V}/d\delta{\bf B} \propto |\delta {\bf B}|^{-\alpha_{P}-1}$). The critical division between the ``patchy'' and ``traditional'' behaviors will then occur at $\alpha_{P} =2$. If the PDF of the magnetic fluctuations that scatter CRs falls more steeply ($\alpha_{P}>2$), then the ``core'' of the PDF dominates CR scattering and the behavior will resemble the traditional model. If the PDF falls more slowly/is more shallow ($\alpha_{P}<2$), then the contributions to CR scattering will be instead dominated by the {\em largest} (non-linear, $\mathcal{O}(1)$) fluctuations in the tails -- our ``patchy'' behavior.

\subsection{Candidate Mechanisms for Intermittent, Small-Scale ISM Structure}
\label{sec:small.physics}

Briefly, we note that both the empirically-observed small-scale scattering structures from Section \ref{sec:small.ism} and the theoretical category of ``intermittent'' structures from Section \ref{sec:intermittency} could be related to a variety of distinct microphysical processes in the ISM. Each of these is, in a sense, a ``candidate'' for the patchy scattering structures. This includes plasma sheets in MHD turbulence \citep{Dong:2022}, turbulent boundary/mixing layers \citep{ji:2019.radiative.turbulent.mixing.layers,yang:2023.radiative.turbulent.mixing.layers}, magnetic mirrors and traps \citep{Chandran:2000,bustard:2020.crs.multiphase.ism.accel.confinement,Lazarian:2021,Tharakkal:2022}, plasmoid instabilities \citep{Fielding:2023}, weak shocks \citep{1973SPhD...18..115K,2020PhRvX..10c1021M,Kempski:2022}, regions with strong dust-CR coupling \citep{squire:2021.dust.cr.confinement.damping,ji:2021.cr.mhd.pic.dust.sims}, and regions where self-confinement has ``run away'' (e.g.\ in ``staircase-like'' instabilities; \citealt{quataert:2021.cr.outflows.diffusion.staircase,huang.davis:2021.cr.staircase.in.outflows,tsung:2021.cr.outflows.staircase}), to name a few. 

CR scattering by rare (i.e. not volume-filling) regions of large field-line curvature, proposed recently by \citet{Lemoine:2023} and \citet{Kempski:2023}, is a promising microphysical origin of the patchy transport model. Small-scale bends of magnetic field lines may be a generic intermittent feature of MHD turbulence and exist even on scales much smaller than the turbulence injection scale. These bends are therefore plausible candidates for the ``patches” in our patchy transport model. The scattering rates derived in \citet{Lemoine:2023} and \citet{Kempski:2023} depend on the volume-filling fractions of the large-curvature patches, broadly consistent with the calculations presented here. In particular, our result that $\ell_{\rm mfp} \sim \ell_{ S}/f_{V}$ is somewhat similar to equation (3.1.) in \citet{Lemoine:2023}.

All of these scenarios appear viable in principle, but none has reached the level of theoretical development where we can plot an unambiguous prediction in Fig.~\ref{fig:fvol}. However, our hope is that in future studies modeling the structures predicted by such mechanisms, Fig.~\ref{fig:fvol} can prove a useful ``figure of merit'' to test whether such mechanisms are (or are not) viable candidates for explaining observed CR scattering in the ISM.

\section{Observational Tests}
\label{sec:obs}

A key property of the patchy CR scattering models discussed here is that they, {\em by definition}, produce the same observational constraints and phenomenologically-derived CR properties (e.g. CR scattering rate $\langle \bar{\nu}\rangle_{\rm eff}$, scattering/deflection mean free path $\ell_{\rm mfp}$, CR residence time, grammage, etc.) as ``traditional'' continuous CR scattering models. So long as the CR residence time (observationally inferred to be $\gtrsim$\,Myr) is longer than the scattering time $\sim 1/\langle \bar{\nu}\rangle_{\rm eff}$ (which for the example observationally-inferred values quoted in Section \ref{sec:patchy} is $\sim 30\,$yr at $\sim 1\,$GeV), or equivalently so long as the total ``path length'' traveled along the trajectory of a given CR is larger than one mean free path (equivalent to saying that the grammage exceeds $X > \rho_{\rm ISM}\,\ell_{\rm mfp} \sim 10^{-5}\,{\rm g\,cm^{-2}}\,R_{\rm GV}^{-0.5}$, which is satisfied by several orders of magnitude), this ensures all of the predictions for e.g.\ CR spectra, primary-to-secondary or radioactive or isotopic ratios, etc., will be identical in a ``patchy'' model and a ``continuous/traditional'' model with the same  $\langle \bar{\nu}\rangle_{\rm eff}$. This means, for example, that while one might naively expect a ``patchy'' model to produce a larger observable CR anisotropy, such a difference would only be present in the population at distances $\lesssim \ell_{\rm mfp} \lesssim 10\,{\rm pc}$ from the initial CR sources, and the CRs will (by definition) converge to isotropic on a timescale $\sim 1/\langle \bar{\nu}\rangle_{\rm eff} \sim \ell_{\rm mfp}/{\rm v}_{\rm cr} \sim 30\,$yr. Since no SNR exists so close to Earth, we predict no difference in the anisotropy of local CRs. 

More detailed tests of e.g.\ higher-order statistics and correlations will, in general, require a specific model for the origin of the ``patchy'' CR scattering (e.g.\ different specific physical scenarios discussed in Section \ref{sec:examples}). However, we can propose some generic predictions that may conceivably be testable with future instruments. 

Consider, for example, a scattering patch of depth $\ell_{S}$ with internal scattering rate $\nu_{s}$, and a ``CR scattering optical depth'' of $\tau_{s} \sim \ell_{s} \nu_{s}/{\rm v}_{\rm cr}$ (Section \ref{sec:constraints:sizes}). By definition, this patch can scatter, i.e. temporarily confine, CRs below some critical rigidity $R_{s}$, enhancing their density relative to the ambient medium by a factor $\sim \tau_{s}$ (just like with multiply-scattered photons). This would, in principle, enhance the $\gamma$-ray emission at some energies from the patch. However, that effect alone would be strictly degenerate with local variations in the source density, variations in the volume-filling scattering rate in the ``traditional'' models, and variations in the background nucleon number density of the ISM. But since, by definition, CRs with rigidities $R\gg R_{s}$ are not confined by the patch, their relative density is not enhanced. Therefore, the patches would exhibit a $\gamma$-ray spectrum that is more biased to emission from $R<R_{s}$ compared to the ambient medium. Not only would such variation exist, but the change in spectral shape would be correlated with the sizes and number densities of the scattering structures as we have discussed above (because we require a spectrum of patches that scatter CRs of different energies with different relative rates). 

The challenge here is that measuring such an effect would require orders-of-magnitude superior resolution in $\gamma$-ray telescopes compared to current state-of-the art instruments like Fermi. Ideally, observing this effect would require the ability to resolve the $\gamma$-ray spectra at $\sim 0.1-100$\,GeV energies with spatial resolution $\sim \ell_{S} \sim$\,au (or at least $\ll 10^{3}$\,au) -- i.e.\ angular resolution in the ISM (for typical spatial distances) reaching sub-arcsecond levels, whereas current instruments typically achieve few-degree resolution at these energies. 

Alternatively, one could look for a similar effect in radio synchrotron emission from CR electrons with similar energies, where the angular resolution of current instruments is much less limited. But even in this case, the desired angular resolution is still a challenge, and far beyond the scope of current single-dish surveys, requiring interferometry with $\gg$\,km baselines. More problematically, the synchrotron spectral slope at these energies is also strongly influenced by the relative strength of inverse Compton and synchrotron losses, which would be at least partially degenerate with the desired measurement. Still, despite these challenges, there may already be hints of the relative/patchy CR enhancement described above in existing observations, perhaps related to inferences of an excess of low-energy ionizing CRs in GMCs \citep[e.g.][]{indriolo:2009.high.cr.ionization.rate.clouds.alt.source.models,indriolo:2012.cr.ionization.rate.vs.cloud.column,indriolo:2015.cr.ionization.rate.vs.galactic.radius,Yang:2014,Baghmanyan:2020}, or to the observed ``point-source-like'' excess of $\sim1$\,GeV (aka ``soft'' according to the arguments above) $\gamma$-ray emission from the Galactic center \citep{hooper:2011.galactic.center.excess,Cholis:2015,lee:2016.galactic.center.excess.point.sources,bartels:2016.galactic.second.excess.pulsar.args,ackermann:2017.fermi.galactic.center.excess,hooper:2017.galactic.center.gamma.ray.excess}.

Another possible set of tests would be to look for the variations caused by discreteness noise in the ``number of scattering structures'' on scales comparable to $\ell_{\rm mfp}$ around specific CR sources. We caution that the point of interest here is not the emission from the acceleration region or CR source itself (e.g.\ not emission from SNRs or PWNe), but the secondary emission from the ISM on $\sim$\,pc scales around such a source. Here one must be careful to estimate out to which radii the CR background ``seen by'' the ISM would be dominated by the local source, versus the collective Galactic background. Moreover, one has the same resolution challenges as noted above, since the most obvious tests would require resolving inhomogeneity on a spatial scale of order $\sim \ell_{S}$.

\section{Summary}
\label{sec:summary}

In light of the theoretical challenges facing ``traditional'' CR scattering theories, which assume low-energy ($\sim$ MeV-TeV) CRs are scattered by a uniform, volume-filling population of weak magnetic field fluctuations, we propose a novel category of ``patchy'' CR scattering models, in which CRs are scattered by inhomogeneous/intermittent/punctuated structures with small volume-filling factors. We derive a number of constraints any such structures must obey (e.g. their sizes, internal CR scattering rates/magnetic field fluctuations, number densities, mass and volume-filling factors) in order to be both internally self-consistent and to reproduce existing observational constraints. We show that fundamentally, in this category of models, we can reproduce the observed dependence between CR scattering and rigidity ($\langle \bar{\nu} \rangle_{\rm eff} \sim R^{-\delta_s}_{\rm GV}$) by imposing a size distribution of scattering structures: larger ``patches'' (with greater optical depth to higher-energy CRs) are rarer but scatter a wider range of CR energies, while smaller ``patches'' are more abundant but only scatter lower-energy CRs. 

We consider a variety of observational and physical candidates for such structures. We show that many ``macroscopic'' quasi-static ISM structures (e.g.\ GMCs, SNRs, PNe, stellar magnetospheres, HII regions, PWNe) cannot be the dominant scattering regions since the mean-free path between them, as seen by CRs, is too large. However we show that small-scale or intermittent structures in the ISM, with size scales as small as $\lesssim$\,au and volume-filling factors as small as $\sim 10^{-7}$, could plausibly explain the observed CR scattering rates from $\sim$\,MeV-TeV energies, while avoiding any obvious theoretical or observational objections. These may even be related to other small-scale ISM structures inferred from radio-wave scattering observations. However, as of yet, there is no single obvious physical mechanism that is clearly predicted to produce the desired scattering rates. We propose a ``figure of merit,'' akin to a ``Hillas plot'' for CR pitch-angle scattering in the ISM, with which to compare any such future models.

We discuss some direct observational tests of this model category, though we conclude that, for now, the required resolution remains far beyond current capabilities.

\acknowledgments{I.S.B. was supported by the DuBridge Postdoctoral Fellowship at Caltech as well as by NASA through Hubble Fellowship grant HST-HF2-51525.001-A awarded by the Space Telescope Science Institute, which is operated by the Association of Universities for Research in Astronomy, Incorporated, under NASA contract NAS 5-26555. Support for P.F.H. was provided by NSF Research Grants 1911233, 20009234, NSF CAREER grant 1455342, NASA grants 80NSSC18K0562, HST-AR-15800.001-A. P.K. was supported by the Lyman Spitzer, Jr. Fellowship at Princeton University. J.S. acknowledges the support of the Royal Society Te Ap\=arangi, through Marsden-Fund grant MFP-UOO2221 and Rutherford Discovery Fellowship  RDF-U001804.  This research was facilitated by the Multimessenger Plasma Physics Center (MPPC), NSF grant PHY-2206610, by a Simons Investigator award to EQ, and by NSF AST grant 2107872.}
\datastatement{The data supporting this article are available on reasonable request to the corresponding author.} 
\bibliography{main}
\end{document}